\documentclass[twocolumn,english,aps,superscriptaddress,amsmath,amssymb,floatfix]{revtex4-1}
\usepackage[colorlinks=true,citecolor=blue,linkcolor=magenta]{hyperref}

\usepackage[markup=blue, authormarkupposition=left]{changes} 

\usepackage{soul}
\usepackage[utf8]{inputenc}
\usepackage[english]{babel}
\usepackage{amsmath,amsfonts,amssymb}
\usepackage[T1]{fontenc}
\usepackage{url}

\usepackage{amsmath}
\usepackage{amsfonts}
\usepackage{amssymb}

\usepackage{epstopdf}
\usepackage{graphicx}

\begin{document}

\title{Tunable X-band opto-electronic synthesizer with ultralow phase noise}

\author{Igor Kudelin}
\affiliation{Electrical Computer \& Energy Engineering, University of Colorado, Boulder, CO 80309, USA}
\affiliation{Department of Physics, University of Colorado Boulder, 440 UCB Boulder, CO 80309, USA}
\affiliation{National Institute of Standards and Technology, 325 Broadway, Boulder, CO 80305, USA}

\author{Pedram Shirmohammadi}
\affiliation{Department of Electrical and Computer Engineering, University of Virginia, Charlottesville, VA 22904, USA}

\author{William Groman}
\affiliation{Electrical Computer \& Energy Engineering, University of Colorado, Boulder, CO 80309, USA}
\affiliation{Department of Physics, University of Colorado Boulder, 440 UCB Boulder, CO 80309, USA}
\affiliation{National Institute of Standards and Technology, 325 Broadway, Boulder, CO 80305, USA}

\author{Samin Hanifi}
\affiliation{Department of Electrical and Computer Engineering, University of Virginia, Charlottesville, VA 22904, USA}

\author{Megan L. Kelleher}
\affiliation{Department of Physics, University of Colorado Boulder, 440 UCB Boulder, CO 80309, USA}
\affiliation{National Institute of Standards and Technology, 325 Broadway, Boulder, CO 80305, USA}

\author{Dahyeon Lee}
\affiliation{Department of Physics, University of Colorado Boulder, 440 UCB Boulder, CO 80309, USA}
\affiliation{National Institute of Standards and Technology, 325 Broadway, Boulder, CO 80305, USA}

\author{Takuma Nakamura}
\affiliation{Department of Physics, University of Colorado Boulder, 440 UCB Boulder, CO 80309, USA}
\affiliation{National Institute of Standards and Technology, 325 Broadway, Boulder, CO 80305, USA}

\author{Charles A. McLemore}
\affiliation{Department of Physics, University of Colorado Boulder, 440 UCB Boulder, CO 80309, USA}
\affiliation{National Institute of Standards and Technology, 325 Broadway, Boulder, CO 80305, USA}

\author{Alexander Lind}
\affiliation{Electrical Computer \& Energy Engineering, University of Colorado, Boulder, CO 80309, USA}
\affiliation{National Institute of Standards and Technology, 325 Broadway, Boulder, CO 80305, USA}

\author{Dylan Meyer}
\affiliation{Electrical Computer \& Energy Engineering, University of Colorado, Boulder, CO 80309, USA}

\author{Junwu Bai}
\affiliation{Department of Electrical and Computer Engineering, University of Virginia, Charlottesville, VA 22904, USA}

\author{Joe C. Campbell}
\affiliation{Department of Electrical and Computer Engineering, University of Virginia, Charlottesville, VA 22904, USA}

\author{Steven M. Bowers}
\affiliation{Department of Electrical and Computer Engineering, University of Virginia, Charlottesville, VA 22904, USA}

\author{Franklyn Quinlan}
\affiliation{Electrical Computer \& Energy Engineering, University of Colorado, Boulder, CO 80309, USA}
\affiliation{National Institute of Standards and Technology, 325 Broadway, Boulder, CO 80305, USA}

\author{Scott A. Diddams}
\affiliation{Electrical Computer \& Energy Engineering, University of Colorado, Boulder, CO 80309, USA}
\affiliation{Department of Physics, University of Colorado Boulder, 440 UCB Boulder, CO 80309, USA}
\affiliation{National Institute of Standards and Technology, 325 Broadway, Boulder, CO 80305, USA}

\begin{abstract}

Modern communication, navigation, and radar systems rely on low noise and frequency-agile microwave sources. In this application space, photonic systems provide an attractive alternative to conventional microwave synthesis by leveraging high spectral purity lasers and optical frequency combs to generate microwaves with exceedingly low phase noise. However, these photonic techniques suffer from a lack of frequency tunability, and also have substantial size, weight, and power requirements that largely limit their use to laboratory settings. In this work, we address these shortcomings with a hybrid opto-electronic approach that combines simplified optical frequency division with direct digital synthesis to produce tunable low-phase-noise microwaves across the entire X-band. This results in exceptional phase noise at 10 GHz of -156 dBc/Hz at 10 kHz offset and fractional frequency instability of 1$\times$10$^{-13}$ at 0.1 s. Spot tuning away from 10 GHz by $\pm$500 MHz, $\pm$1 GHz, and $\pm$2 GHz, yields phase noise at 10 kHz offset of -150 dBc/Hz, -146 dBc/Hz, and -140 dBc/Hz, respectively. The synthesizer architecture is fully compatible with integrated photonic implementations that will enable a versatile microwave source in a chip-scale package. Together, these advances illustrate an impactful and practical synthesis technique that shares the combined benefits of low timing noise provided by photonics and the frequency agility of established digital synthesis.
\end{abstract}

\maketitle

\section*{Introduction}

Microwave signals with low phase and timing noise are critical for multiple fields of wide scientific, technological, and societal impact. This includes the areas of precision timekeeping, navigation, communications and radar-based sensing. Increasing demands for improved performance in these applications drive the need for microwave frequency synthesis that provides both low timing jitter and broad tunability. For example, applications like microwave spectroscopy at the core of atomic clocks~\cite{townes2013microwave}, as well as radio astronomy~\cite{sullivan2012classics} 
require precise and low-noise signals, but with minimal frequency tuning. On the other hand, communication systems often require fast frequency hopping, where the phase noise on the microwave carrier affects the error vector magnitude of the signal and bit error rate~\cite{khanzadi2015phase, razavi2012rf, armada2001understanding}. Radar is another prominent application that relies on both frequency agility and low noise~\cite{serafino2019toward, ghelfi2014fully, mckinney2014photonics, skolnik2001introduction}. Here, low phase noise improves the detection probability as well as the imaging accuracy and quality~\cite{krieger2006impact}. Beyond these examples, low-noise and frequency-agile synthesis is indispensable in other technical and scientific fields including metrology, sensors, and navigation systems. The extent of the application space elevates the importance of robust and low noise microwave synthesis that covers a broad frequency range.

Fundamentally, there exists a trade off in frequency synthesis between low noise and broad and fast tunability. Conventional electronic synthesizers introduce tunability with multiple electronic up-converters and local oscillators. These tuning elements provide an avenue to introduce noise that is multiplicative as the frequency increases. In contrast, the best photonic-based synthesis employs frequency division to deliver microwaves with unprecedented noise performance~\cite{Diddams2003, fortier2011generation, xie2017photonic, nakamura2020coherent, eliyahu2008phase}; however, this performance comes with restricted tunability that is often in the range of only 0.1\%. In this work, we challenge this paradigm by combining the ultralow noise of photonic-based frequency division with broadband electronic digital tuning. In contrast to previous works~\cite{fortier2016optically, wei2018all}, our advance incorporates significant simplifications to the photonic system that are compatible with recent advances in chip-level integration~\cite{Kudelin2024, li2014electro, Sun2024, Zhao2024, jin2024microresonatorreferenced, Li:23, he2024chip}. 

Central to the architecture of our synthesizer is a low-noise photonic oscillator at 10 GHz that is generated using 2-point optical frequency division (2P-OFD)~\cite{swann2011microwave, li2014electro, Kudelin2024}. Compared to traditional OFD approaches~\cite{Diddams2003, fortier2011generation,fortier2016optically}, 2P-OFD results in a lower division factor, but enables a significant reduction in the size and power requirements. The photonic part of our system provides a fixed 10 GHz microwave signal with phase noise of -156 dBc/Hz at 10 kHz offset and fractional frequency instability of 1$\times$10$^{-13}$. This low-noise signal then serves as the reference clock for a direct digital synthesizer (DDS), the output of which is mixed with the clock itself to provide tunable low noise microwaves across the entire X-band (8-12 GHz). This yields microwaves with phase noise at 10 kHz offset of -150 dBc/Hz, -146 dBc/Hz, and -140 dBc/Hz in the tuning range of $\pm$500 MHz, $\pm$1 GHz, and $\pm$2 GHz from the 10 GHz, respectively. At the same time, the DDS allows tuning with $\mu$Hz resolution and speeds of tens of ns. Compared to previous works~\cite{fortier2016optically, wei2018all}, we increase the continuous tuning range by up to a factor of 4, with improvement in the phase noise of 10 dB. Importantly, this is accomplished with a simplified system amenable to chip-scale integration to address the demanding requirements of modern applications beyond the laboratory environment.


\begin{figure}[!htb]
\centering
\includegraphics[width=\linewidth]{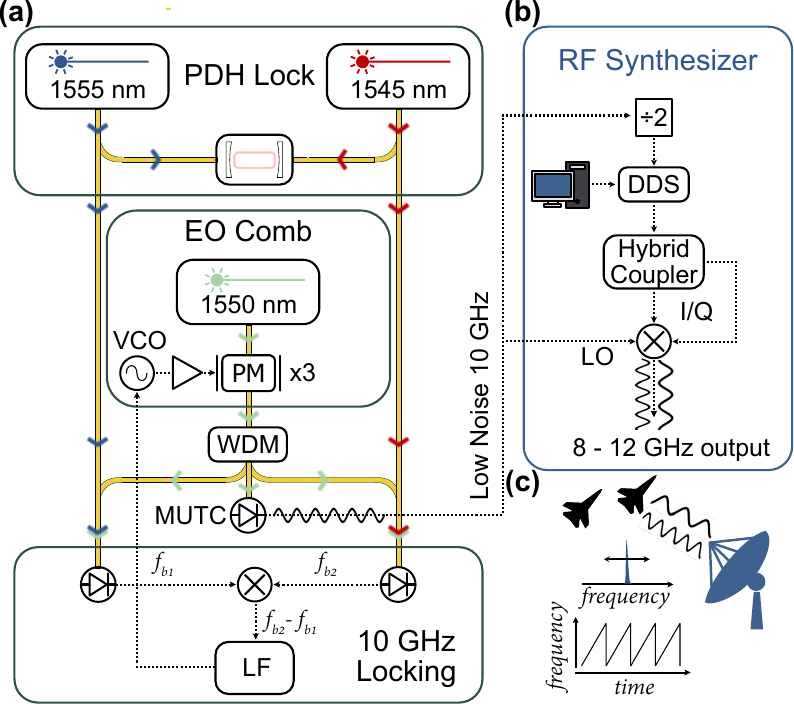}
\caption{\textbf{Experimental setup. (a)} Two CW fiber lasers at 1545 nm and 1555 nm are locked to a millimeter-scale Fabry-P\'{e}rot (FP) cavity with the Pound-Drever-Hall (PDH) technique. The comb is generated by modulating the light from a 1550 nm CW laser at 10 GHz. The EO comb is split in a WDM to provide the beat notes ($f_{b1}$ and $f_{b2}$) with the reference lasers. Part of the EO comb is detected in a modified uni-traveling carrier (MUTC) detector to provide a low noise microwave at $f_{rep}$. The beat frequencies between the CW lasers and the EO comb are mixed together to provide the error signal that is conditioned with a loop filter (LF) and used to stabilized the VCO and the EO comb mode spacing. \textbf{(b)} The stabilized 10 GHz is frequency-divided by two and serves as the clock signal for a direct digital synthesizer (DDS). The DDS output is split in the hybrid coupler to provide two signals with 90\textdegree ~relative phase shift. These signals are mixed with the original 10 GHz in an IQ mixer for single-sideband generation. By tuning the the DDS frequency, the synthesizer output covers the entire X-band (8 - 12 GHz). \textbf{(c)} Illustration of applications that rely on low-noise frequency sweeping, such as radar and navigation systems. }
\label{fig:f1}
\end{figure}

\section*{Low-noise Microwave Generation}

In our realization of 2P-OFD, the low phase noise of continuous wave (CW) lasers is transferred to an optical frequency comb and its microwave-rate mode spacing. This is achieved by first narrowing the linewidth of the CW lasers through active stabilization of their frequencies to a high quality (Q) factor Fabry-P\'{e}rot (FP) cavity~\cite{drever1983laser}. The FP cavity plays a crucial role in low-noise microwave generation with 2P-OFD by providing the phase and frequency reference of the generated microwave signal. The lowest noise can be achieved with a long FP cavity length~\cite{hafner20158, alvarez2019optical} or with cryogenic cooling~\cite{matei2017}. However, to reduce the system size, we take advantage of recent advances in miniature FP cavities \cite{McLemore2022Mini, kelleher2023compact} and employ an FP with 6.3 mm cavity length and 23.6 GHz free-spectral range (FSR). Ultralow loss cavity mirrors yield a cavity with Q$\sim$5 billion and the fractional frequency stability as low as $3\times 10^{-14}$ with 1 s of averaging~\cite{kelleher2023compact}. The phase noise of the lasers that are stabilized to the cavity inherit the thermal-limited cavity length stability for offset frequencies below 10 kHz. 

\begin{figure}[!tb]
\centering
\includegraphics[width=\linewidth]{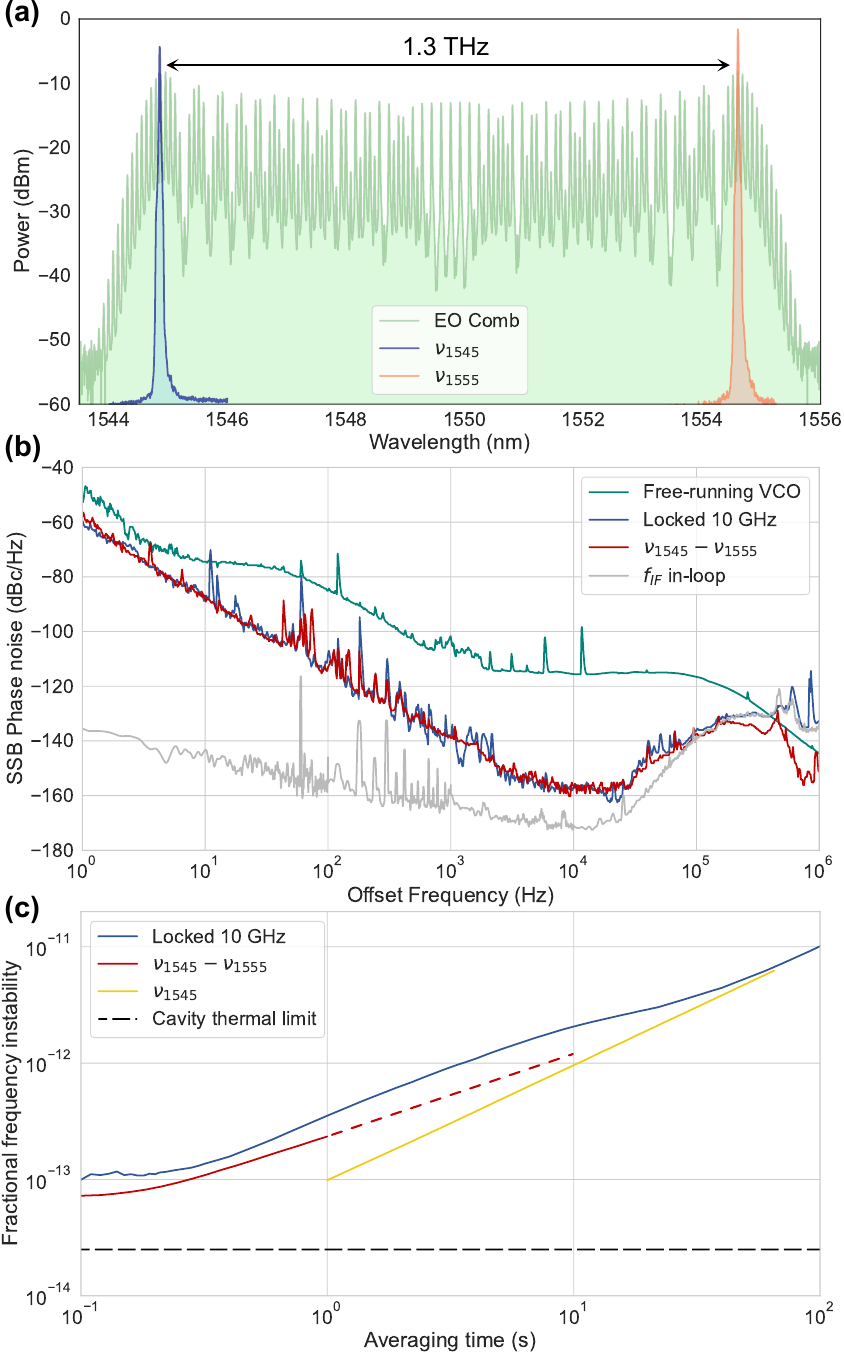}
\caption{\textbf{Electro-optic comb performance. (a)} Optical spectra of the EO comb and reference CW lasers. \textbf{(b)} The phase noise of the free-running VCO (yellow), photodetected stabilized 10 GHz (blue), in-loop phase noise of the intermediate frequency locking (light grey), and the relative phase noise of the reference CW lasers ($\nu_{1545} - \nu_{1555}$) spaced by 1.3 THz. This noise is measured independently and decreased by the OFD value of 42 dB (red). \textbf{(c)} Fractional frequency instability of the stabilizied 10 GHz (blue), relative stability of reference CW lasers ($\nu_{1545} - \nu_{1555}$) spaced by 1.3 THz (red), the optical stability of the reference laser (yellow), and the calculated thermal noise limit of the cavity (dotted black). 
}
\label{fig2:f2}
\end{figure}

As shown in the experimental setup in Fig.~\ref{fig:f1}, two lasers with frequencies $\nu_{1555}$ and $\nu_{1545}$ and a frequency separation of 1.3~THz  are stabilized to different modes of the FP cavity. This gap, which is $55\times$ the cavity FSR, is then divided down with an electro-optic (EO) frequency comb. The EO comb is generated by phase modulating a 1550 nm CW laser ($\nu_{1550}$) with the amplified output of a voltage-controlled oscillator (VCO) at $f_m=10$ GHz \cite{parriaux2020electro} (see Supplemental Data for full details). The comb frequency spacing is given by $f_{rep}=f_m$ and the optical spectrum of the comb along with the two CW lasers is shown in Fig.~\ref{fig2:f2}(a).

2P-OFD is implemented by heterodyning the cavity-stabilized optical references $\nu_{1555}$ and $\nu_{1545}$ with the closest comb lines $\nu_{1550} - n f_{rep}$ and $\nu_{1550} + m f_{rep}$, where $n$ and $m$ are positive integers. This yields two beat-notes $f_{b1} = \nu_{1555} - \nu_{1550} + n f_{rep}$ and $f_{b2} =  \nu_{1545} - \nu_{1550} - m f_{rep}$, which are further mixed to provide an intermediate frequency (IF) $f_{IF} = f_{b2} - f_{b1} = \nu_{1545} - \nu_{1555} - (n + m)f_{rep}$. The correct choice of the sum or difference between $f_{b1}$ and $f_{b2}$ depends on the frequency positions of the CW lasers relative to the comb lines. 
Conveniently, $f_{IF}$ does not depend on the center frequency of the EO comb, allowing for the use of a free-running laser for $\nu_{1550}$. To complete the comb stabilization, the intermediate frequency is compared to a reference oscillator $f_{ref}$ to generate an error signal that is conditioned and fed to the 10 GHz VCO. 

Once the servo loop is closed, the frequency of $f_{rep}$ is given by:
\begin{equation}
    f_{\rm rep} = \frac{(\nu_{\rm 1545} - \nu_{\rm 1555}) + f_{\rm IF}}{(n+m)}.
\label{eq:frep}
\end{equation}
The denominator $(n+m)$ is the number of comb modes between the CW lasers and is responsible for the frequency division and corresponding noise reduction due to OFD. In terms of the phase noise power spectral density, this reduction is equal to $20\log [(\nu_{1545} - \nu_{1555})/f_{rep}] = 20\log[1.3~\rm THz/10 ~\rm GHz] = 42$ dB. This division reduces the noise contributions of the relative stability of the reference lasers, $(\nu_{1545} - \nu_{1555})$, and $f_{IF}$.

The output of the servo-controlled VCO provides direct access to the 10 GHz signal. In principle, this signal could be at the power level of several Watts if taken at the termination port of one of the modulators (See Supplementary materials). Since we use three pairs of RF amplifiers and EOMs in parallel, the control feedback loop for comb stabilization accounts for all additive noise of the RF amplifiers. However, the 10 GHz signal taken after just one of the amplifiers would not have the noise of the other RF amplifiers suppressed by the servo loop. That would limit the achievable noise to be greater than that of the reference lasers $(\nu_{1545} - \nu_{1555})$ (see Supplementary Materials). Instead, to suppress the noise of all of the RF amplifiers and obtain the lowest-noise 10 GHz signal, we photodetect the output of the EO comb with a high-power and high-linearity modified uni-traveling carrier (MUTC) photodiode~\cite{Fortier2013Photonic,zang2018reduction}. Fiber dispersion and optical filtering of the spectrum in the WDM transforms the phase modulation that creates the EO comb into an amplitude modulated signal that can be photodetected with sufficient signal-to-noise ratio.

As shown in Fig.~\ref{fig2:f2}(b), in this configuration we achieve phase noise of -156 dBc/Hz at 10 kHz offset frequency (Fig.~\ref{fig2:f2}(b)). At higher offset frequencies the microwave performance is limited by the phase locking of $f_{IF}$ (grey curve of Fig.~\ref{fig2:f2}(b)). Reducing the noise in this frequency range would require broader bandwidth servo control of $f_{IF}$. At offset frequencies below 100 kHz, the phase noise is limited by the relative stability of the CW lasers, as shown by the red curve of Fig.~\ref{fig2:f2}(b). Since the reference lasers are PDH locked to the common cavity, they both follow the thermal noise of the cavity and exhibit noise correlation, leading to improved relative phase noise that can be below thermal noise limit of the cavity~\cite{liu2023cmr}. This common mode rejection of the cavity noise reduces the phase noise to the limit imposed by residual noise of the individual laser locking circuits.

We also characterize the time-domain stability of the 10 GHz microwave signal, with results shown in Fig.~\ref{fig2:f2}(c). Here we measure a minimum in the fractional frequency instability of 1$\times$10$^{-13}$ at integration time of 0.1~s. At longer averaging time, the instability increases due to the drift of the FP cavity. This drift is mirrored by the drift of a single CW laser as well as the relative instability of the two CW lasers. However, we note that the frequency instability between the two CW lasers is higher than the stability of each of the reference CW laser. We attribute that to the measurement setup (see Supplementary materials) that is susceptible to environmental vibrations. All of the curves of Figure~\ref{fig2:f2}(c) are above the previously measured instability limit of the cavity which is at the level of 2-3$\times$10$^{-14}$\cite{kelleher2023compact}.

As indicated in Eq.\ref{eq:frep} and surrounding text, $f_{rep}$ can be tuned via $f_{\rm IF}$, which assumes a maximum value of $f_{rep}$ through coarse stepwise tuning of the VCO. Nonetheless, when divided by $(n+m)\approx130$ the tunability is reduced to a small fraction of $f_{rep}$. This limitation is a common drawback in all OFD systems, but in what follows we show how the tuning range can be significantly increased. 

\begin{figure*}[!ht]
\centering
\includegraphics[width=\linewidth]{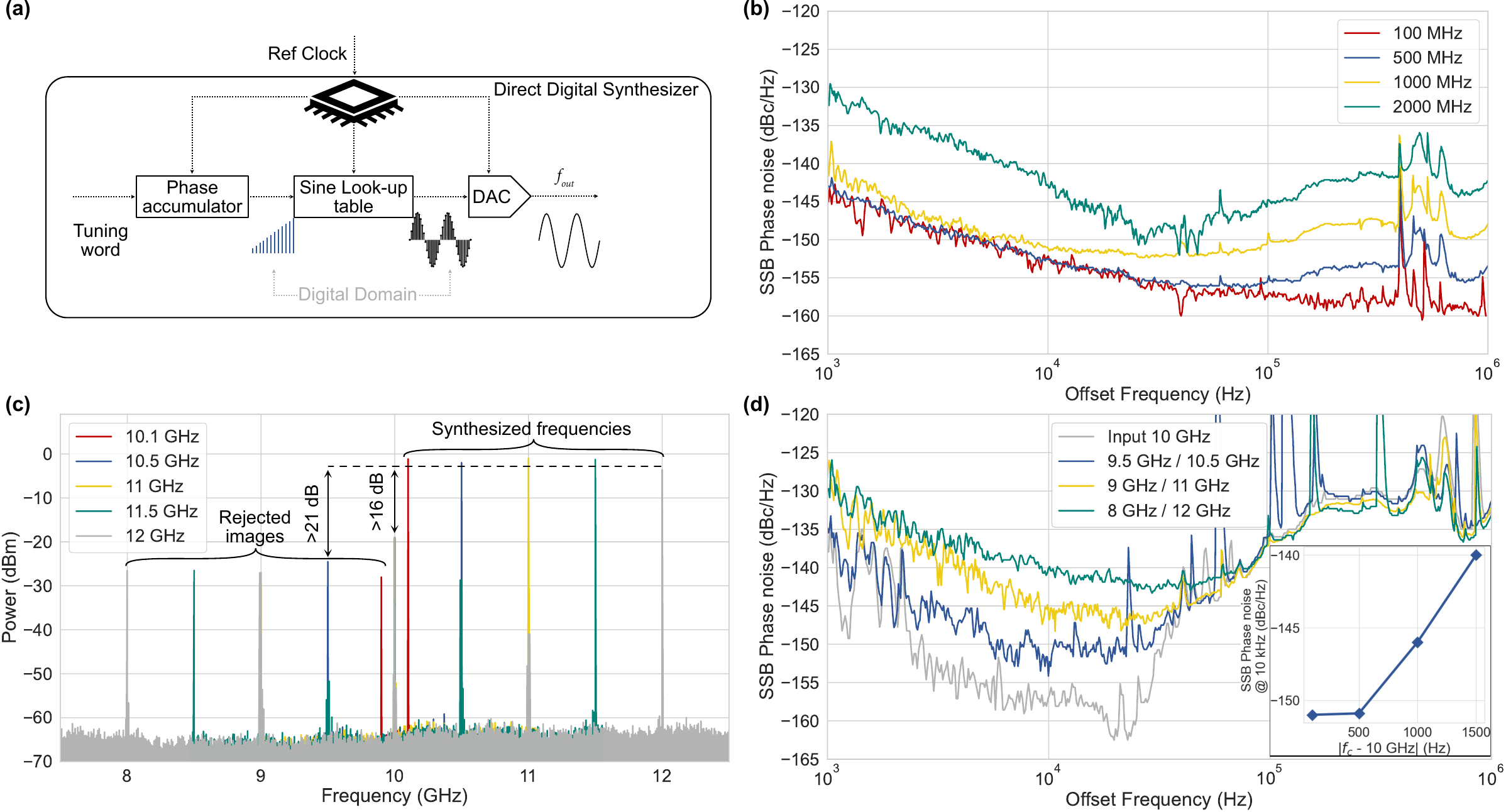}
\caption{\textbf{Synthesizer performance.} \textbf{(a)} DDS architecture. \textbf{(b)} Single side-band (SSB) phase noise of DDS output in the range from 100 MHz to 2 GHz, while the stabilized 10 GHz serves as an external clock. \textbf{(c)} RF spectra of the synthesized frequencies. The spectra with lower syntehesized frequencies is exactly symmetric to the presented one. \textbf{(d)} Single side-band (SSB) phase noise of the synthesized frequencies from 8 GHz to 12 GHz. Inset: Phase noise at 10 kHz offset against the synthesized frequency. 
}
\label{fig:f3}
\end{figure*}

\section*{Broad Bandwidth Tunability with Low Phase Noise}

Broad bandwidth tunability of the low-noise 10 GHz microwave is achieved by mixing this signal with the output of a direct digital synthesizer (DDS). With reference to Fig.~\ref{fig:f1}(b), the low noise microwave at 10 GHz is frequency divided by 2 and serves as a reference (clock) for the DDS. The DDS output is then added to (or subtracted from) the original 10 GHz reference using an IQ mixer. 
Since the DDS initially creates the output waveform only at discrete time intervals, its generated output frequency can not exceed the Nyquist–Shannon sampling theorem limit. In the case of the DDS employed here, the maximum output frequency is 2 GHz, which is $\sim$40\% of the reference input $f_{rep}/2$.

The basic architecture of the DDS is shown in Figure~\ref{fig:f3}(a). 
At each clock cycle, the phase accumulator calculates a new phase value by taking into account both the clock signal frequency and a user-provided tuning word. It then performs a lookup operation to translate this phase information into a digital amplitude that is transformed into an analog voltage via the digital-to-analogue converter (DAC)~\cite{all_dds}. 
The two main contributions to the phase noise of the DDS output are: (1) the input clock noise that is reduced by the ratio $(f_{clock} / f_{out})^2$, and (2) the intrinsic noise originating within the DDS itself, which arises from  quantization noise, truncation noise and nonlinearity in the DAC~\cite{dds_phase}.

In Fig.~\ref{fig:f3}(b) we show sample phase noise when the DDS is clocked by the low-noise 5 GHz and its output is set at frequencies between 100 MHz and 2 GHz. For the 100 MHz output, the output noise is limited by the DDS intrinsic noise, reaching -152 dBc/Hz at 10 kHz offset frequency. As the DDS frequency increases, the noise of the clock signal starts to dominate at offset frequencies above 100 kHz.


The DDS output is summed or differenced with original 10 GHz microwave in an IQ mixer. 
The low noise 10 GHz at $f_{rep}$ is amplified to saturate the mixer, while the DDS output is split in a hybrid coupler to provide a 90\textdegree~phase shift between the I and Q ports of the mixer. The phase shift of 90\textdegree~is necessary to achieve single sideband operation with high image rejection. Output in the range of 8-10 GHz or 10-12 GHz is determined by the relative sign of the phase shift between the I and Q ports. 

Power spectra of synthesized frequencies in the 10-12 GHz band are shown in Fig.~\ref{fig:f3}(c). The rejection of the synthesized frequencies relative to the original 10 GHz carrier is more than 16 dB across all tuning frequencies. This rejection can be increased with higher power driving the I and Q inputs of the mixer. At the same time, the image rejection exceeds 21 dB across all generated frequencies. This rejection can be improved with more precise amplitude balance and phase control between the I and Q ports. Alternatively, one could use two DDSs, clocked by the same microwave input, with precise digital tuning of the amplitude and respective phases adjusted to 90\textdegree.

Figure~\ref{fig:f3}(d) shows our measurements of the phase noise of the synthesized frequencies at discrete values across the entire X-band. At offsets above 80 kHz, the phase noise is limited by that of the 10 GHz signal itself, while at lower offset frequencies the noise from the DDS dominates. The noise on the synthesized frequencies in the range of 9.5 to 10.5 GHz falls below -150 dBc/Hz at 10 kHz offset, while the phase noise increases to near -140 dBc/Hz at 8 and 12 GHz. 

This increase versus carrier frequency ($f_c$) is summarized in the inset in Fig.~\ref{fig:f3}(d). The phase noise contribution of the DDS to the entire synthesizer that can be categorized in two regions. In the range of 9.5 to 10.5 GHz, the phase noise does not depend on carrier frequency. However, at carrier frequencies offset by more than 500 MHz from 10 GHz, the phase noise increases approximately as $20\log(N)$, where $N$ is the frequency multiplication factor. Considering the noise sources described earlier, we assume that the dominant noise source at carrier frequencies closer to 10 GHz is quantization noise in the digital-to-analog conversion. This noise results from the quantized number of DAC bits and the symbol rate of the DAC. Meanwhile, the noise originating within the final output stage of DAC is likely the limitation at higher frequencies. This noise arises from both the DAC switching mechanism and the flicker noise of the output stages from the analog circuits, which is scaled by generated frequency. We note that since the noise is primarily limited by the DDS, there is no need for a reference microwave source with noise lower than the internal noise of DDS, making 2P-OFD an ideal technique for such a synthesis approach.

An advantage of our approach is its versatility that overcomes the obstacle of limited tunability of photonic-based microwave generators. We expect that similar phase noise performance can readily be achieved in other microwave bands. For example, by using the 20 GHz harmonic from the photodetected EO comb, it is possible to partially cover the K-band while maintaining low noise, similar to that of Figure~\ref{fig:f3}(d). The tuning resolution of the DDS is in the $\mu$Hz range, and that can be controlled at the speeds of tens of ns~\cite{wei2018all}. Compared to the previous work on tunable microwave generation with frequency comb~\cite{fortier2016optically, wei2018all}, we demonstrate 4-fold improvement in the tunability with up to 10 dB improvement in phase noise. 

\section*{Discussion \& Summary} 

In this work we demonstrate tunable low noise microwave generation using 2P-OFD in combination with DDS. To the best of our knowledge, this work provides the first 10 GHz microwave generation via 2P-OFD with low noise on both short and longer timescales. This is demonstrated via frequency instability of 1$\times$10$^{-13}$ at 0.1 s and the phase noise of -156 dBc/Hz at 10 kHz offset. This milestone is significant because it is realized with components amenable to low-SWaP on-chip microwave generation, where 2P-OFD is the ideal approach in terms of simplicity and performance~\cite{Kudelin2024, li2014electro, Sun2024, Zhao2024, jin2024microresonatorreferenced, he2024chip}. Another significant milestone of the work is the first demonstration of broad frequency tunability with 2P-OFD, while maintaining low phase noise. Key to this is using the stabilized 10 GHz signal as the clock of the DDS. 
For the synthesized frequencies in the range of 8 GHz to 12 GHz the system supports the phase noise of -140 dBc/Hz at 10 kHz offset. And in the range of 9.5 GHz to 10.5 GHz, the phase noise is below -150 dBc/Hz. In comparison to other similar works~\cite{fortier2016optically, wei2018all}, we showed a substantial improvement in the tuning range with up to 10 dB improvement in the phase noise. 
While this work demonstrated the advantages of combining 2P-OFD with DDS-based synthesis, the performance still can be improved by optimizing the system with lower noise DDS and lasers, larger OFD factors, and improved servo systems. Additionally, with lower noise RF amplifiers it is possible to access low noise microwave with power above +30 dBm (see Supplementary Materials). These insights provide a roadmap for low-SWaP photonic microwave generation with low-noise and broad tunability, as will be important for multiple applications in navigation, communications and sensing.

\bibliographystyle{ieeetr}

\medskip
\begin{footnotesize}

\noindent \textbf{Corresponding authors}: \href{mailto:igor.kudelin@colorado.edu}{igor.kudelin@colorado.edu} and \href{mailto:scott.diddams@colorado.edu}{scott.diddams@colorado.edu}

\noindent \textbf{Funding}: 
This research was supported by DARPA GRYPHON program (HR0011-22-2-0009), NIST and the University of Colorado.

\noindent \textbf{Acknowledgments}: 
The authors thank K. Chang and N. Hoghooghi for comments on the manuscript. Commercial equipment and trade names are identified for scientific clarity only and does not represent an endorsement by NIST.

\noindent \textbf{Author Contributions}:

F.Q. and S.A.D. conceived the experiment and supervised the project.
I.K., and S.A.D. wrote the paper with input from all authors.
I.K., A.L., and W.G., built the experiment setup.
I.K., and D.M. performed the experiments.
M.L.K., and F.Q. built the F-P cavity.
P.S., S.H., I.K., and S.M.B. built the RF synthesizer.
D.L., T.N., C.A.M., and F.Q. provided the optically derived microwave reference and aided in the microwave phase noise measurement system.
J.B., and J.C.C. provided MUTC detectors.
All authors contributed to the system design and discussion of the results.

\noindent \textbf{Disclosures}: 

The authors declare no conflicts of interest.

\noindent \textbf{Data and materials availability}:  

All data can be requested from the corresponding authors.

\end{footnotesize}

\end{document}